\documentstyle[twocolumn,floats,aps,epsf]{revtex}
\begin{document}
\twocolumn[
\hsize\textwidth\columnwidth\hsize\csname@twocolumnfalse%
\endcsname
\draft
\title{Phase diagrams of period-4 spin chains 
consisting of three kinds of spins}
\author{Ken'ichi Takano \\
Laboratory of Theoretical Condensed Matter Physics and 
Research Center for Advanced Photon Technology, \\
Toyota Technological Institute, Nagoya 468-8511, Japan}
\date{Received \today}
\maketitle
\begin{abstract}
      We study a period-4 antiferromagnetic mixed quantum 
spin chain consisting of three kinds of spins. 
      When the ground state is singlet, the spin magnitudes 
in a unit cell are arrayed as $(s-t, s, s+t, s)$ with integer 
or half-odd integer $s$ and $t$ ($0 \le t < s$). 
      The spin Hamiltonian is mapped onto a nonlinear $\sigma$ 
model (NLSM) in a previously developed method. 
      The resultant NLSM includes only two independent 
parameters originating from four exchange constants 
for fixed $s$ and $t$. 
      The topological angle in the NLSM determines the gapless 
phase boundaries  between disordered phases in the parameter 
space. 
      The phase diagrams for various $s$ and $t$ shows rich 
structures. 
      We systematically explain the phases 
in the singlet-cluster-solid picture. 
\end{abstract}
\pacs{PACS numbers: 75.10.Jm, 75.30.Kz}
]

\narrowtext

\section{Introduction}

      The lowest spin excitation for a homogeneous 
antiferromagnetic spin chain has a gap if the spin magnitude 
is an integer, and has no gap if it is a half odd integer. 
      Haldane conjectured this proposition by 
mapping the spin chain onto the nonlinear $\sigma$ model 
(NLSM) \cite{Haldane}. 
      From the topological term of the NLSM, we know whether 
the spin excitation is gapless or not. 
      Mapping to the NLSM is now recognized as one of powerful 
methods to examine quantum spin systems. 
      Various aspects of the NLSM method 
are found in Refs. \cite{Affleck,Fradkin,Tsvelik}. 

      Analyses using the NLSM have been extended to 
inhomogeneous spin chains. 
      The first step has been done by Affleck 
\cite{Affleck2,Affleck3}. 
      He reformulated the NLSM method in an operator formalism 
in which spin operators are transformed in pairs. 
      Due to the pair transformation, his formalism is applicable 
to a spin chain with bond alternation. 
      Then the exchange constant is inhomogeneous with period 2 
but the spin magnitude is still homogeneous. 
      Results by the NLSM method qualitatively agree with 
numerical and experimental results 
\cite{Kato,Yamamoto,Hagiwara}. 

      The NLSM method for a general inhomogeneous spin chain 
with arbitrary period is given, if the ground state is singlet, 
in a previous paper \cite{Takano1}. 
      The inhomogeneity is not only for the exchange constant 
but also for the spin magnitude. 
      The derivation of the NLSM is based on dividing the spin 
chain into blocks and simultaneously transforming the spin 
variables belonging to a block in a path integral formalism. 
      From the topological term, the gapless condition for 
the spin excitation is obtained as an equation, which 
we have called the {\it gapless equation}. 
      The gapless equation determines the phase boundaries 
in the ground-state phase diagram. 
      This method is quite general, and is applicable to various 
spin chains. 

      A period-4 chain is the simplest nontrivial case in which 
more than one kind of spins can be mixed under condition that 
the ground state is singlet. 
      We have applied the NLSM method to a period-4 
chain consisting of two kinds of spins and presented 
the phase diagram in the parameter space of exchange constants 
\cite{Takano2,Takano3}. 
      When the magnitudes of two kinds of spins are $s_a$ and 
$s_b$,  only the array of ($s_a$, $s_a$, $s_b$, $s_b$) 
in a unit cell consists with a singlet ground state.  
      We have constructed an NLSM and then the gapless 
equation for this case following the general procedure 
\cite{Takano1}; 
      Fukui and Kawakami also derived the 
NLSM in a different standpoint \cite{Fukui1}. 
      We constructed phase diagrams from the NLSM and found 
that they generally consist of many disordered phases. 
      To understand various phases we have proposed the 
singlet-cluster-solid (SCS) picture, which is an extension of 
the valence-bond-solid (VBS) picture \cite{Affleck4}. 
      The SCS picture systematically explains all phases for any 
values of $s_a$ and $s_b$. 
      A simple version of the SCS picture is seen in the case of 
$s_a$=$s_b$=$\frac{1}{2}$ \cite{Chen1}. 
      In the case of $s_a$=$\frac{1}{2}$ and $s_b$=1, numerical 
calculation has been performed \cite{Hikihara1}. 
      The phase diagram by the NLSM method qualitatively 
agrees with the numerical phase diagram.  

      In this paper, we study a period-4 spin chain consisting of 
three kinds of spins. 
      Systematic treatment of this problem seems to be difficult 
due to many possibilities in the choice of 
the spin kinds. 
      However, the condition that the ground state is singlet 
fairly restricts allowed combinations. 
      In fact, the spin 
magnitudes in a unit cell must be generally arrayed as 
$(s-t, s, s+t, s)$ with integer or half-odd integer $s$ 
and $t$ ($0 \le t < s$), as will be seen. 
      The spin chain has a modulation in the spin magnitude 
around {\it average} $s$; 
      hence it is interesting to compare results for different 
$t$ and the same $s$. 
      The exchange constants are also periodic with period 4 
and then there are three parameters except for an energy unit. 
      The spin Hamiltonian is mapped onto an NLSM in the 
general method \cite{Takano1}. 
      The resultant NLSM shows that the number of relevant 
parameters is not three but two. 
      The phase diagrams for various $s$ and $t$ determined 
by the gapless equation show rich phase structures 
\cite{Takano4}. 
      We systematically explain the phases in the SCS picture. 

      This paper is organized as follows. 
      In Sec.~II, we introduce the Hamiltonian for the period-4 
spin chain consisting of three kinds of spins with a singlet 
ground state, and parameterize the exchange constants. 
      In Sec.~III, the spin Hamiltonian is transformed to an NLSM. 
      From the topological term of the NLSM, 
the gapless equation to determine phase boundaries is derived. 
      In Sec.~IV, phase diagrams are drawn by the gapless 
equation for various spin magnitudes. 
      Features of the phases are mentioned. 
      In Sec.~V, the ground states of the phases are explained 
by the SCS picture. 
      Section VI is devoted to summary and discussion.

\section{Hamiltonian} 

     The spin Hamiltonian with period 4 is generally written as 
\begin{eqnarray}
\label{Hamiltonian}
      H = \sum_{j=1}^{N/4} &{}& 
 ( J_1 \, {\bf S}_{4j+1} \cdot {\bf S}_{4j+2} 
 + J_2 \, {\bf S}_{4j+2} \cdot {\bf S}_{4j+3} 
\nonumber \\
 &{}&+ J_3 \, {\bf S}_{4j+3} \cdot {\bf S}_{4j+4} 
 + J_4 \, {\bf S}_{4j+4} \cdot {\bf S}_{4j+5} ) , 
\end{eqnarray}
where ${\bf S}_j$ is the spin at site $j$ with magnitude $s_j$. 
      The number of lattice sites is $N$, the lattice spacing 
is $a$ and the system size is $L = a N$. 
      We only consider the antiferromagnetic exchange interaction 
($J_i > 0$). 

      Since the system is periodic with period 4 in the spin 
magnitude as well as in the exchange constant, we generally 
have 4 different spin magnitudes 
($s_1$, $s_2$, $s_3$, $s_4$) in a unit cell. 
      A unit cell is illustrated in Fig.~\ref{unit_cell}. 
      The values of the spin magnitudes are restricted 
in order that the system has a singlet ground state. 
      Following the Lieb-Mattis theorem \cite{Lieb},  
a singlet ground state is realized if the system 
satisfies the restriction 
\begin{equation}
\label{restriction}
       s_1 - s_2 + s_3 - s_4 = 0 ; 
\end{equation}
otherwise the system has a ferrimagnetic ground state. 
       We have treated the case that three kinds of spins 
are mixed. 
       To consist with the restriction (\ref{restriction}), 
the spin magnitudes must be of the following form: 
\begin{eqnarray}
\label{magnitude}
(s_1, s_2, s_3, s_4) = (s-t, s, s+t, s) , 
\end{eqnarray}
where $s$ and $t$ are positive integers or half-odd integers 
satisfying $s>t \ge 0$. 
       The spin configuration (\ref{magnitude}) is regarded as 
a modulation against the uniform configuration 
$(s, s, s, s)$. 

       We parameterize the exchange constants as 
\begin{eqnarray}
\label{exchange}
&{}& J_{1} = \frac{J}{1+\delta} , \quad 
      J_{2} = \frac{J'}{1-\gamma} , 
\nonumber \\
&{}& J_{3} = \frac{J'}{1+\gamma} , \quad 
      J_{4} = \frac{J}{1-\delta} , 
\end{eqnarray}
where $\delta$ ($\gamma$) is the distortion parameter 
describing the asymmetry between exchange constants 
$J_4$ and $J_1$ ($J_2$ and $J_3$) 
on the both sides of a spin with $s_1$ ($s_3$). 
\begin{figure}[btp]
\begin{center}\leavevmode
\epsfxsize=50mm
\epsfbox{unit_cell.EPSF}
\caption{Spin magnitudes and exchange constants 
of the Hamiltonian (1) with 
Eqs. (3) and (4) in a unit cell.}
\label{unit_cell}
\end{center}
\end{figure}

\section{Mapping to the nonlinear $\sigma$ model}

      The expectation value of a spin operator in a coherent state is
expressed as 
\begin{equation}
           <{\bf S}_j> = (-1)^j s_j {\bf n}_j 
\label{coherent}
\end{equation}
with a unit vector ${\bf n}_j$. 
      The partition function $Z$ at temperature $1/\beta$ is 
then represented as 
\begin{eqnarray}
      Z &=& \int \! D[{\bf n}_j] \, 
\prod_j \delta({\bf n}^2_j - 1) \, e^{-S} , 
\label{partition}
\\
      S &=& i \sum_{j=1}^{N} (-1)^j s_j w[{\bf n}_j] 
\nonumber \\ 
         &{}& + \frac{1}{2} 
\int_0^{\beta} \!\! d\tau \sum_{j=1}^{N} 
                J_j s_j s_{j+1} ({\bf n}_j - {\bf n}_{j+1})^2 . 
\label{action_S} 
\end{eqnarray}
      In the action (\ref{action_S}), the first term comes from the 
Berry phase and 
$w[{\bf n}_j]$ is the solid angle which the unit  vector ${\bf n}_j$
forms in the period $\beta$. 

      We have derived the NLSM action starting from the action 
(\ref{action_S}) for the general periodic case \cite{Takano1}. 
      The derivation is based on dividing the spin chain into 
blocks and transforming the spin variables in a block into new 
ones. 
      In the present case, by choosing unit cells as blocks, 
the transformation for the $p$th block is written as 
\begin{eqnarray}
{\bf n}_{4p+1} &=& \frac{3}{4}{\bf m}(p) 
+ \frac{1}{4}{\bf m}(p-1) + a {\bf L}_1(p) , 
\label{p_1}
\nonumber \\ 
{\bf n}_{4p+2} &=& \ \ \, {\bf m}(p) 
\qquad \qquad \quad \ \ \, + a {\bf L}_2(p) , 
\label{p_2}
\nonumber \\ 
{\bf n}_{4p+3} &=& \frac{3}{4}{\bf m}(p) 
+ \frac{1}{4}{\bf m}(p+1) + a {\bf L}_3(p) , 
\label{p_3}
\nonumber \\ 
{\bf n}_{4p+4} &=& \frac{1}{2}{\bf m}(p) 
+ \frac{1}{2}{\bf m}(p+1) + a {\bf L}_4(p) , 
\label{p_4}
\end{eqnarray}
where \{${\bf m}(p)$\} are gradually changing unit vectors 
and \{${\bf L}_q(p)$\} are small fluctuations. 
       This transformation does not change the number of the original 
degrees of freedom \cite{Takano1}. 
      Integrating out the fluctuations $\{ {\bf L}_q(p) \}$ and 
taking the continuum limit, we obtain the effective action 
\cite{Takano5}: 
\begin{eqnarray}
&{}&      S_{\rm eff} = \int^{\beta}_0 \!\! d\tau \int^{L}_0 \!\! dx 
\biggl\{ 
      - i \frac{J^{(0)}}{J^{(1)}} 
{\bf m} \cdot (\partial_{\tau} {\bf m} \times \partial_x {\bf m}) 
\nonumber \\ 
&{}&      + \frac{1}{2aJ^{(1)}} \biggl( 
\frac{J^{(1)}}{J^{(2)}} - \frac{J^{(0)}}{J^{(1)}} \biggl) 
(\partial_{\tau} {\bf m})^2 
      + \frac{a}{2} J^{(0)} (\partial_x {\bf m})^2 
\biggl\} , 
\label{action-NLSM}
\end{eqnarray}
where
\begin{eqnarray}
\frac{1}{J^{(0)}} &=& \frac{1}{2s} \left[ \frac{1}{J (s-t)} 
+ \frac{1}{J '(s+t)}  \right] , 
\label{def_J_0} 
\\
\frac{1}{J^{(1)}} &=& \frac{s-t}{4s} 
\left[ \frac{1}{J(s-t)} + \frac{1}{J'(s+t)} 
+ \frac{d}{J'(s-t)} \right] , 
\label{def_J_1}
\\
\frac{1}{J^{(2)}} &=& \frac{1}{4s(s+t)} 
\left( \frac{s^2-t^2}{J} + \frac{s^2+t^2}{J'} 
+ \frac{s^2-t^2}{J'}d \right) . 
\label{def_J_2}
\end{eqnarray}
      We have used the effective distortion parameter 
\begin{equation}
d \equiv \gamma + \delta \, \frac{J'}{J} . 
\label{distortion_parameter}
\end{equation}
      It is remarkable that two distortion parameters 
$\gamma$ and $\delta$ appear only in a single parameter $d$. 
      Hence the system is characterized by only a pair of 
parameters, $(J'/J, d)$. 
      The value of $d$ is restricted as 
\begin{equation}
| d | < 1 + \frac{J'}{J} , 
\label{restriction_distortion}
\end{equation}
because $| \delta | < 1$ and $| \gamma | < 1$ from 
Eq.~(\ref{exchange}). 

      The action (\ref{action-NLSM}) is of the standard 
form of the NLSM: 
\begin{eqnarray}
       S_{\rm st} &=& \int^{\beta}_0 \!\! d\tau \int^{L}_0 \!\! dx 
\biggl\{ 
- i \frac{\theta }{4\pi} 
{\bf m} \cdot (\partial_{\tau} {\bf m} \times \partial_x {\bf m}) 
\nonumber \\ 
      &+& \frac{1}{2gv} (\partial_{\tau} {\bf m})^2 
+ \frac{v}{2g} (\partial_x {\bf m})^2 
\biggl\} 
\label{standard_NLSM}
\end{eqnarray}
with the topological angle $\theta$, the coupling constant $g$ 
and the spin wave velocity $v$. 
      Comparing Eq.~(\ref{action-NLSM}) to 
Eq.~(\ref{standard_NLSM}), $\theta$ is given as 
\begin{eqnarray}
\label{angle}
      \theta = 4\pi \frac{J^{(0)}}{J^{(1)}} . 
\end{eqnarray}

\section{Phase diagrams} 

      The NLSM has a gapless excitation 
when $\theta/2\pi$ is a half-odd integer. 
      From Eq.~(\ref{angle}), 
this condition is rewritten as the gapless equation:  
\begin{equation}
\frac{1}{J^{(1)}} = \frac{2l_0-1}{4} \frac{1}{J^{(0)}} , 
\label{gapless_condition}
\end{equation}
where $l_0$ is an integer. 
       For each $l_0$, this equation determines a 
boundary between disordered phases if it has a solution. 

      Substituting Eqs.~(\ref{def_J_0}) and (\ref{def_J_1}) 
into Eq.~(\ref{gapless_condition}), we have  
\begin{equation}
d  = \left( l - s - t - \frac{1}{2} \right) 
\left( \frac{1}{s+t} + \frac{1}{s-t} \frac{J '}{J} \right) 
\label{gapless_equation}
\end{equation}
with $l = l_0 + 2t + 1$.
      A phase boundary is a straight line in the parameter 
space of $(J'/J, d)$. 
       Allowed values of integer $l$ satisfying 
Eq.~(\ref{restriction_distortion}) are 
\begin{equation}
l = 1, 2, \cdots, 2(s+t) . 
\label{allowed_l}
\end{equation}
      Thus we have found that there are $2(s+t)+1$ phases 
separated by $2(s+t)$ gapless phase boundaries. 
      All of them go through the point 
\begin{equation}
\left( - \frac{s-t}{s+t}, 0 \right). 
\label{focus}
\end{equation}

      In the special case of $d$ = 0, Eq.~(\ref{gapless_equation}) 
reduces to $l = s + t + \frac{1}{2}$. 
      Then, the system has a spin gap when $s_3$ (= $s + t$) 
is an integer, while it has no spin gap when $s_3$ is 
a half-odd integer. 
      This is an extended version of Haldane's proposition for 
$t$ = 0 and $J'$ = $J$ to cases with period-4 modulation 
with respect to the spin magnitude and the exchange interaction. 
      The condition $d$ = 0 (i.~e. $J \gamma + J' \delta$ = 0) 
is satisfied even for nonzero $\gamma$ and $\delta$. 
      This means that effects of distortions can be canceled. 

\begin{figure}[btp]
\begin{center}\leavevmode
\epsfxsize=70mm
\epsfbox{Phase_10_00.EPSF}

\leavevmode
\epsfxsize=70mm
\epsfbox{Phase_10_05.EPSF}
\end{center}
\caption{The phase diagrams of $s$ = 1 for (a) $t$ = 0, 
and (b) $\frac{1}{2}$. 
      Unshaded regions are physical.}
\label{Phase_10}
\end{figure}
      In Fig.~\ref{Phase_10}, we present phase diagrams for 
$s$ = 1. 
      Figure ~\ref{Phase_10}(a) is for $s$ = 1 and $t$ = 0; 
spins are arrayed as 1-1-1-1 in a unit cell. 
      Then the spin magnitude is homogeneous, while 
the exchange interaction is modulated with period 4. 
      The shaded regions are not physical because of 
Eq.~(\ref{restriction_distortion}) and $J'/J$ $>$ 0. 
      There are three phases labeled by $A_+$, $H$, and $A_-$. 
      Phase $H$ is the Haldane phase since it includes the point 
(1, 0) which represents the uniform $s$ = 1 spin chain. 
      Two phases $A_+$ and $A_-$ appear, when the effective 
distortion is strong. 
      Figure ~\ref{Phase_10}(b) is for $s$ = 1 and 
$t$ = $\frac{1}{2}$; 
      spins are arrayed as $\frac{1}{2}$-1-$\frac{3}{2}$-1 
in a unit cell. 
      There are four phases labeled by $B_+$, $A_+$, $A_-$, 
and $B_-$. 
      In contrast to the case of (a), the system is on a boundary 
when there is no effective distortion ($d$=0). 
      It is noticed that the phases $B_+$ and $B_-$ appear, 
when $J'/J$ $<$ 1. 
      All the phases in Figs.~\ref{Phase_10}(a) and (b) 
are interpreted by means of the SCS picture 
in the next section. 

\begin{figure}[btp]
\begin{center}\leavevmode
\epsfxsize=70mm
\epsfbox{Phase_15_00.EPSF}

\leavevmode
\epsfxsize=70mm
\epsfbox{Phase_15_05.EPSF}

\leavevmode
\epsfxsize=70mm
\epsfbox{Phase_15_10.EPSF}
\caption{The phase diagrams of $s$ = $\frac{3}{2}$ for 
(a) $t$ = 0, (b) $\frac{1}{2}$, and (c) 1. 
      Unshaded regions are physical.}
\label{Phase_15}
\end{center}
\end{figure}

\begin{figure}[btp]
\begin{center}\leavevmode
\epsfxsize=70mm
\epsfbox{Phase_20_00.EPSF}

\leavevmode
\epsfxsize=70mm
\epsfbox{Phase_20_05.EPSF}

\leavevmode
\epsfxsize=70mm
\epsfbox{Phase_20_10.EPSF}

\leavevmode
\epsfxsize=70mm
\epsfbox{Phase_20_15.EPSF}
\end{center}
\caption{The phase diagrams of $s$ = 2 for 
(a) $t$ = 0, (b) $\frac{1}{2}$, (c) 1, and (d) $\frac{3}{2}$. 
      Unshaded regions are physical.}
\label{Phase_20}
\end{figure}

      In Fig.~\ref{Phase_15}, we present phase diagrams for 
$s$=$\frac{3}{2}$. 
      Figure ~\ref{Phase_15}(a) is for $s$=$\frac{3}{2}$ and 
$t$=0;  spins are arrayed as 
$\frac{3}{2}$-$\frac{3}{2}$-$\frac{3}{2}$-$\frac{3}{2}$
in a unit cell. 
      Only the exchange interaction is modulated with period 4. 
      There are 4 phases labeled by $B_+$, $A_+$, $A_-$, 
and $B_-$. 
      These phases, including $B_+$ and $B_-$ for strong distortion, 
develop from small $J'/J$ to large $J'/J$. 
      Figure ~\ref{Phase_15}(b) is for $s$=$\frac{3}{2}$ and 
$t$=$\frac{1}{2}$; 
      spins are arrayed as 1-$\frac{3}{2}$-2-$\frac{3}{2}$ 
in a unit cell. 
      There are 5 phases labeled by $B_+$, $A_+$, $H$, $A_-$, 
and $B_-$. 
      Phases $B_+$ and $B_-$ appear only when $J'/J < 1$. 
      Figure ~\ref{Phase_15}(c) is for $s$=$\frac{3}{2}$ and 
$t$=1;  spins are arrayed as 
$\frac{1}{2}$-$\frac{3}{2}$-$\frac{5}{2}$-$\frac{3}{2}$
in a unit cell. 
      There are 6 phases labeled by $C_+$, $B_+$, $A_+$, $A_-$, 
$B_-$, and $C_-$. 
      Phases $C_+$, $B_+$, $B_-$, and $C_-$ appear 
only when $J'/J < 1$; 
      $C_+$ and $C_-$ are for strong distortion and for 
very small $J'/J$. 
      In the case of (a), the system is on a gapless boundary 
when there is no effective distortion ($d$=0). 
      All the phases in Figs.~\ref{Phase_15}(a), (b) and (c) 
are interpreted by means of the SCS picture 
in the next section. 

      In Fig.~\ref{Phase_20}, we present phase diagrams for 
$s$=2. 
      Figures 4(a), (b), (c) and (d) are for $t$=0, $t$=$\frac{1}{2}$, 
$t$=1, and $t$=$\frac{3}{2}$, respectively; 
spins are arrayed as 2-2-2-2, $\frac{3}{2}$-2-$\frac{5}{2}$-2, 
1-2-3-2, and $\frac{1}{2}$-2-$\frac{7}{2}$-2 in a unit cell. 
      Phase $H$ for small $d$ in Figs. 4(a) and (c) is the $s$=2 
Haldane phase, which includes the point (1, 0). 
      In the cases of (b) and (d), the system is on a gapless boundary 
when there is no effective distortion ($d$=0). 
      Some phases develop from small $J'/J$ to large $J'/J$, 
and the others are restricted in the region $J'/J < 1$.

\section{Singlet-cluster-solid picture} 

      Disordered phases in the phase diagrams are explained 
in the SCS picture, which includes the VBS picture as a 
special case. 
      In the SCS picture, a spin with more than $\frac{1}{2}$ 
magnitude is decomposed into $\frac{1}{2}$ spins. 
      The original spin state is retrieved by symmetrizing 
the states of the decomposed $\frac{1}{2}$ spins at each site. 
      A disordered state is represented by a regular array of 
local singlet clusters of even numbers of 
$\frac{1}{2}$ spins, while a state in the VBS picture is 
a direct product of only singlet dimers. 

\begin{figure}[btp]
\begin{center}\leavevmode
\epsfxsize=60mm
\epsfbox{SCS_10_00.EPSF}
\end{center}
\caption{
      The SCS pictures for the phases of $s$=1 and $t$=0. 
      They represent states for phases (a) $A_+$, (c) $H$, 
and (e) $A_-$ in Fig. 2(a). 
      $H$ is the Haldane phase. 
      They are the same as the VBS pictures in the present case. 
      Gapless states on the phase boundaries are presented in 
(b), and (d). 
}
\label{SCS_10_00}
\end{figure}
\begin{figure}[btp]
\begin{center}\leavevmode
\epsfxsize=60mm
\epsfbox{SCS_10_05.EPSF}
\end{center}
\caption{
      The SCS pictures for the phases of 
$s$=1 and $t$=$\frac{1}{2}$. 
      They represent states for phases (a) $B_+$, (c) $A_+$, 
(e) $A_-$, and (g) $B_-$ in Fig. 2(b). 
      Gapless states on the phase boundaries are presented in 
(b), (d), and (f). 
}
\label{SCS_10_05}
\end{figure}
      The SCS pictures for the phase diagrams of $s$=1 in 
Fig.~\ref{Phase_10} are shown in Figs.~\ref{SCS_10_00} 
and \ref{SCS_10_05}. 
      Figure~\ref{SCS_10_00} explains the phase diagram in 
Fig.~\ref{Phase_10}(a) for the 1-1-1-1 spin chain 
($s$=1, $t$=0). 
      For each picture, a small circle represents a $\frac{1}{2}$ 
spin; an original $s$=1 spin is decomposed into two 
$\frac{1}{2}$ spins. 
      A loop represents a singlet dimer of two $\frac{1}{2}$ 
spins in it. 
      A dashed line represents a spatially extended singlet 
state. 
      Picture (a) ((e)) is for a state in the dimer phase $A_+$ 
($A_-$), and has advantage for the exchange energy 
on $J_2$ and/or $J_4$ ($J_1$ and/or $J_3$) interactions 
because of positive (negative) $d$. 
      Picture (c) is for a translationally invariant state in 
the Haldane phase $H$. 
      These pictures are nothing but the VBS picture; 
a ground state is represented only by singlet dimers. 
      Pictures (b) and (d) are for states between $A_+$ and $H$, 
and between $H$ and $A_-$, respectively; 
      each picture includes an extended singlet state 
(dashed line) contributing to a gapless excitation. 

      Figure~\ref{SCS_10_05} explains the phase diagram in 
Fig.~\ref{Phase_10}(b) for the 
$\frac{1}{2}$-1-$\frac{3}{2}$-1 spin chain 
($s$=1, $t$=$\frac{1}{2}$). 
      Pictures (c) and (e) are for dimer states 
$A_+$ and $A_-$, respectively. 
      The state (c) ((e)) has advantage for the exchange energy 
on $J_2$ and/or $J_4$ ($J_1$ and/or $J_3$) interactions 
because of positive (negative) distortion $d$. 
      Pictures (a) and (g) are for singlet cluster states 
in the phases $B_+$ and $B_-$, which are  
in the region of $J'/J < 1$. 
      In phase $B_+$ ($B_-$), a singlet cluster or dimer including 
a $J_3$ ($J_2$) interaction is the most unfavorable for 
the exchange energy because of large $|d|$ and small $J'/J$. 
      Hence clusters of six $\frac{1}{2}$ spins are formed 
to avoid singlet dimers at $J_3$ ($J_2$) interactions. 
      Pictures (b), (d), and (f) are for gapless states 
on the phase boundaries, where extended states appear. 

\begin{figure}[btp]
\begin{center}\leavevmode
\epsfxsize=60mm
\epsfbox{SCS_15_00.EPSF}
\end{center}
\caption{
      The SCS pictures for the phases of 
$s$=$\frac{3}{2}$ and $t$=0. 
      They represent states for phases (a) $B_+$, (b) $A_+$, 
(c) $A_-$, and (d) $B_-$ in Fig. 3(a). 
      They are the same as the VBS pictures in the present case. 
}
\label{SCS_15_00}
\end{figure}
\begin{figure}[btp]
\begin{center}\leavevmode
\epsfxsize=60mm
\epsfbox{SCS_15_05.EPSF}
\end{center}
\caption{
      The SCS pictures for the phases of 
$s$=$\frac{3}{2}$ and $t$=$\frac{1}{2}$. 
      They represent states for phases (a) $B_+$, (b) $A_+$, 
(c) $H$, (d) $A_-$, and (e) $B_-$ in Fig. 3(b). 
}
\label{SCS_15_05}
\end{figure}
\begin{figure}[btp]
\begin{center}\leavevmode
\epsfxsize=60mm
\epsfbox{SCS_15_10.EPSF}
\end{center}
\caption{
      The SCS pictures for the phases of 
$s$=$\frac{3}{2}$ and $t$=1. 
      They represent states for phases (a) $C_+$, (b) $B_+$, 
(c) $A_+$, (d) $A_-$, (e) $B_+$, and (f) $C_-$ in 
Fig. 3(c). 
}
\label{SCS_15_10}
\end{figure}
      The SCS pictures for the phase diagrams of 
$s$=$\frac{3}{2}$ in Fig.~\ref{Phase_15} are shown in 
Figs.~\ref{SCS_15_00}, \ref{SCS_15_05}, and 
\ref{SCS_15_10}. 
      Those for phase boundaries are not drawn to reduce 
the figure sizes. 
      Figure~\ref{SCS_15_00} explains the phase diagram in 
Fig.~\ref{Phase_15}(a) for the 
$\frac{3}{2}$-$\frac{3}{2}$-$\frac{3}{2}$-$\frac{3}{2}$ 
spin chain ($s$=$\frac{3}{2}$, $t$=0). 
      The SCS pictures represent the ground states in phases 
(a) $B_+$, (b) $A_+$, (c) $A_-$, and (d) $B_-$, and are 
the same as the VBS pictures. 
      Picture (a) ((d)) is for the dimer phase $B_+$ ($B_-$), 
where the energy reduction on $J_2$ and/or $J_4$ 
($J_1$ and/or $J_3$) interactions is the most favorable 
because of large $|d|$. 
      Picture (b) ((c)) is for the phase $A_+$ ($A_-$), 
where the energy reduction on $J_2$ and/or $J_4$ 
($J_1$ and/or $J_3$) interactions is favorable to some extent 
because of smaller but finite $|d|$. 

      Figure~\ref{SCS_15_05} explains the phase diagram in 
Fig.~\ref{Phase_15}(b) for the 
1-$\frac{3}{2}$-2-$\frac{3}{2}$ spin chain 
($s$=$\frac{3}{2}$, $t$=$\frac{1}{2}$). 
      The SCS pictures represent states in phases 
(a) $B_+$, (b) $A_+$, (c) $H$, (d) $A_-$, and (e) $B_-$. 
      Picture (c) represents a Haldane-like state. 
      With transitions (c) to (b) and (b) to (a), the number of 
dimers including $J_3$ interactions decreases. 
      This explains that 
$d$ in $B_+$ is larger than $d$ in $A_+$, and 
$d$ in $A_+$ is larger than $d$ in $H$ 
as seen in Fig.~\ref{Phase_15}(b). 
      In particular, the SCS pictures in $B_+$ and $B_-$ 
include singlet clusters of six $\frac{1}{2}$ spins. 
      The large clusters are formed for $J'/J < 1$ to avoid 
singlet dimers including $J_3$ ($J_2$) interactions. 

      Figure~\ref{SCS_15_10} explains the phase diagram in 
Fig.~\ref{Phase_15}(a) for the 
$\frac{1}{2}$-$\frac{3}{2}$-$\frac{5}{2}$-$\frac{3}{2}$ 
spin chain ($s$=$\frac{3}{2}$, $t$=1). 
      The SCS pictures represent states in phases 
(a) $C_+$, (b) $B_+$, (c) $A_+$, (d) $A_-$, (e) $B_-$, 
and (f) $C_-$. 
      Large singlet clusters of more than two $\frac{1}{2}$ 
spins are formed in $C_+$, $B_+$, $B_-$, and $C_-$ 
for $J'/J < 1$ again. 

      The SCS pictures of the ground states for arbitrary $s$ 
and $t$, including the cases of $s$=2 in Fig.~\ref{Phase_20}, 
are similar to the above examples. 
      Phases for small values of $|d|$ are represented by VBS 
pictures, each of which consists of singlet dimers only. 
      The number of possible VBS pictures is $2(s-t)$, and they 
are between the boundaries of $l$=$2t$ and of $l$=$2s+1$ 
in Eqs.~(\ref{gapless_equation}) and (\ref{allowed_l}). 
      For larger $d$ outside the VBS phases in the ($J'/J$, $d$) 
space, there appear phases represented by SCS pictures 
which includes singlet clusters larger than dimers. 
      Generally, as one moves from a phase to another 
accompanied by increasing (decreasing) $d$ for $d>0$ 
($d<0$), the number of singlet dimers on $J_3$ ($J_2$) 
interactions in the SCS picture decreases by one per unit cell. 
      This explains the total number $2(s+t)+1$ of the phases. 
      The phases explained by SCS pictures including singlet 
clusters consisting of more than two $\frac{1}{2}$ spins 
are restricted to small $J'/J$ regions. 
      In fact, singlet dimers including $J_3$ ($J_2$) 
interactions are energetically unfavorable for small $J'/J$,  
and the number of them can be reduced for $t \ne 0$ 
if large clusters not including $J_3$ ($J_2$) interactions 
are formed.

\section{Summary and discussion}

      We studied a period-4 antiferromagnetic mixed quantum 
spin chain consisting of three kinds of spins in the case 
that the ground state is singlet. 
      The spin magnitudes in a unit cell must be arrayed as 
$(s-t, s, s+t, s)$ with integer or half-odd integer $s$ and 
$t$ ($0 \le t < s$). 
      The exchange constants in a unit cell are 
$(J_1, J_2, J_3, J_4 )$; i.~e. 
there are three parameters except for an energy unit. 
      The spin Hamiltonian is transformed to an NLSM 
by a previously developed method \cite{Takano1}. 
      In the NLSM, we find that the number of 
relevant parameters is not three but two. 
      One of them is a ratio $J'/J$ of the exchange energies 
without distortion, 
and the other is the effective distortion parameter $d$. 
      By the gapless equation from the NLSM, 
we determined $2s+2t$ phase boundaries between 
gapful disordered phases in the ($J'/J$, $d$) space 
for each pair of $s$ and $t$. 
      We explained systematically the disordered phases  
by means of the SCS pictures. 
      In the case of $d=0$, the ground state is in a gapful 
Haldane-like phase for $s+t$ being an integer, 
and is on a gapless phase boundary 
for $s+t$ being a half-odd integer. 
      When $d$ increases (decreases) from $d = 0$, 
singlet clusters including $J_3$ ($J_2$) interactions 
successively decreases with phase transitions. 
      If $t \ne 0$, singlet clusters larger than dimers appear 
for large $|d|$ and small $J'/J$. 

      We discuss whether a change from an SCS state to another 
is a phase transition or not. 
      For simplicity, we consider a part of a full SCS state 
as shown in Fig.~\ref{SCS_sym}. 
      State (a) is a VBS state which is a direct product of 
singlet dimers. 
      This state can gradually change to state (b) without 
a phase transition, because state (b) is formed by local 
modifications where a pair of dimers puts together into 
a singlet cluster of four spins. 
      The change from (a) to (c) is similar. 
      We have used picture (a) as a representative of 
(a), (b) and (c) in this paper. 
      On the other hand, there is no way to locally modify 
dimers in (a) to form dimers in (d). 
      That is, the change from (a) to (d) must be realized only 
by a global recombination of dimers or a phase transition. 
      The wave function for (a) is symmetric but that 
for (d) is antisymmetric with respect to the spatial reflection 
about the vertical dotted line. 
      An extended state in (e) appears under the transition 
where both the dimer states are collapsed. 

\begin{figure}[btp]
\begin{center}\leavevmode
\epsfxsize=60mm
\epsfbox{SCS_sym.EPSF}
\end{center}
\caption{
      A part of an SCS picture. 
      Reflection with respect to the dotted vertical line 
is drawn to examine the necessity of a phase transition 
(see text). 
}
\label{SCS_sym}
\end{figure}

      Finally it is expected that materials realizing period-4 
quantum spin chains will be synthesized and experimentally 
studied. 
      The present paper (and Ref. \cite{Takano3}) will hopefully 
work as a guide to investigate such materials.

\section*{Acknowledgment}

      I thank Takashi Tonegawa and Kiyomi Okamoto for 
discussions on the $\frac{1}{2}$-1-$\frac{3}{2}$-1 
spin chain. 
      This work is supported by the Grant-in-Aid for 
Scientific Research from the Ministry of Education, Science, 
Sports and Culture, Japan. 



\begin{references}
\bibitem{Takano5} 
      The NSLM (\ref{action-NLSM}) stands for a general periodic 
spin chain, if general forms are used for $J^{(i)}$ ($i$=0, 1, 2)
\cite{Takano1}.
\end{references}
\end{document}